\documentclass[preprint2,natbib209]{aastex}
\usepackage{natbib}
\newcommand\kms{km\,s$^{-1}$}
\newcommand\hash{$-$}
\newcommand\rich{R$ =$ }
\newcommand\dist{D$ =$ }
\newcommand\h{$^h$}

\newcommand\ad{$^{\circ}$}
\newcommand\po{Paper I}
\begin{document}
\title{Redshifts and Velocity Dispersions of Galaxy Clusters in the
  Horologium-Reticulum Supercluster}   

\author{Matthew C. Fleenor, James A. Rose, Wayne A. Christiansen}
\affil{Department of Physics \& Astronomy, University of North Carolina,
    Chapel Hill, NC 27599}
\email{fleenor2@physics.unc.edu, jim@physics.unc.edu, wayne@physics.unc.edu}

\author{Melanie Johnston-Hollitt}
\affil{Department of Physics, University of Tasmania, TAS 7005, Australia}
\email{Melanie.JohnstonHollitt@utas.edu.au}

\author{Richard W. Hunstead}
\affil{School of Physics, University of Sydney, NSW 2006, Australia}
\email{rwh@physics.usyd.edu.au}

\author{Michael J. Drinkwater}
\affil{Department of Physics, University of Queensland, QLD 4072, Australia}
\email{mjd@physics.uq.edu.au}

\and

\author{William Saunders}
\affil{Anglo-Australian Observatory, Epping NSW 1710, Australia }
\email{will@aaoepp.aao.gov.au}

\begin{abstract}
We present 118 new optical redshifts for galaxies in 12 clusters in the        
Horologium-Reticulum supercluster (HRS) of galaxies.  For 76 galaxies,
the data were obtained with the Dual Beam Spectrograph on the
2.3m telescope of the Australian National University at Siding
Spring Observatory.  After combining 42 previously unpublished
redshifts with our new sample, we determine mean redshifts and
velocity dispersions for 13 clusters, in which previous observational data
were sparse.  In six of the 13 clusters, the newly determined mean 
redshifts differ by more than 750 \kms\ from the published
values.  In the case of three clusters, A3047, A3109, and A3120, the 
redshift data indicate the presence of multiple components along the line 
of sight. The new cluster redshifts, when combined with other reliable
mean redshifts for clusters in the HRS, are found to be distinctly
bi-modal. Furthermore, the two redshift components are consistent with
the bi-modal redshift distribution found for the inter-cluster
galaxies in the HRS by \citet{fle05}.    
\end{abstract}


\section{INTRODUCTION}
The Horologium-Reticulum supercluster (HRS) is an extended region of high      
galaxy density \citep{sha35,luc83,ein03,fle05}, covering $\sim$150 square
degrees of sky at a mean redshift of $\sim$20,000 \kms. The  
HRS also contains more than 20 galaxy clusters \citep{ein97,ein02}. As
discussed in \citet{hud99} and \citet{ein01}, the HRS is the
second largest mass concentration within $\sim$300 Mpc, where it is
only surpassed by the Shapley supercluster (SSC). 
                                                                               
While the SSC has been extensively studied
\citep{qui95,qui00,dri99,dri04,bar98,bar00}, the HRS remains relatively  
unexplored.  Due to the potential importance of such a large-scale 
structure in the present-epoch universe, we have embarked on a redshift
survey to provide a comprehensive mapping of the HRS. Our initial
results, which contain 547 galaxy redshifts in the {\it inter-cluster}
regions of the HRS, are reported in \citet[][hereafter Paper I]{fle05}.
A key result from Paper I is that the distribution of inter-cluster
galaxies is separated into two distinct redshift components.  On
the other hand, the published mean redshifts for 21 galaxy clusters in 
the HRS do not exhibit such a bi-modal distribution. The
differing results between the cluster and inter-cluster redshift
distributions appear to contradict the view that galaxy clusters share
the kinematics of the inter-cluster galaxy distribution as a result of
their location at intersecting filaments of galaxies
\citep[e.g.,][]{van93,bon96,col99,col05}. However, the mean redshift
for many of these clusters is based on fewer than four galaxy
redshifts per cluster, i.e., sparse information. To clarify the
distribution of cluster redshifts in the HRS, we have obtained new
data for 12 clusters in which the previously published data were
sparse.  The results of this program are reported below and, when
combined with other previous redshift data, give an improved
assessment of the distribution of cluster redshifts in the
HRS. Throughout the paper, we adopt the following cosmological parameters:
$\Omega _m = 0.3$, $\Omega _\Lambda = 0.7$, and $H_o = 70 $ \kms\
Mpc$^{-1}$, which implies a spatial scale of 4.6 Mpc degree$^{-1}$ (77 kpc
arcmin$^{-1}$) at the $\sim$20,000 \kms\ mean redshift of the HRS. 

\section{CLUSTER SAMPLE}
Lists of galaxy clusters in the region of the HRS have been taken from
two major studies. The first is the Abell catalog (extension)
\citep[hereafter ACO in][]{aco89}, while the second is the 
Automated Plate Measuring Machine cluster catalog \citep[hereafter
APMCC in][]{dal94,dal97}. Since galaxy clusters represent the largest (at
least partly) virialized structures, they serve as massive signposts
for identifying and studying superclusters of galaxies. Based on the
ACO, \citet{zuc93} identified 18 HRS clusters 
using a combination of partial redshift information and percolation
algorithms.  While working with the same list of ACO clusters,
\citet{ein94} identified 26 members of the HRS. In \po, we used the
17 ACO clusters ocurring in both studies to define
the mean redshift of the HRS ($\overline{cz} =$ 19,900 \kms), and
we adopted the FWHM of the cluster redshift distribution as defining the
HRS kinematic core to lie between 17,000 and 22,500 \kms\ (see Figure 4,
\po). However, the mean redshifts are uncertain for 10 of the 17
ACO clusters because they are based on fewer than four galaxy redshifts each 
\citep[``$N_{\rm{gx}}< 4$'' in][hereafter SR99]{str99}. In this paper we
report new spectroscopic observations, together with previously unpublished
redshifts, for 9 of these 10 clusters with the aim of determining a
more accurate mean redshift and dispersion for each cluster. Published
data for the tenth cluster, A3109, have been reassessed, and
additional spectra have been obtained for a further three clusters
with sparse data in the literature.

\begin{figure}
  \plotone{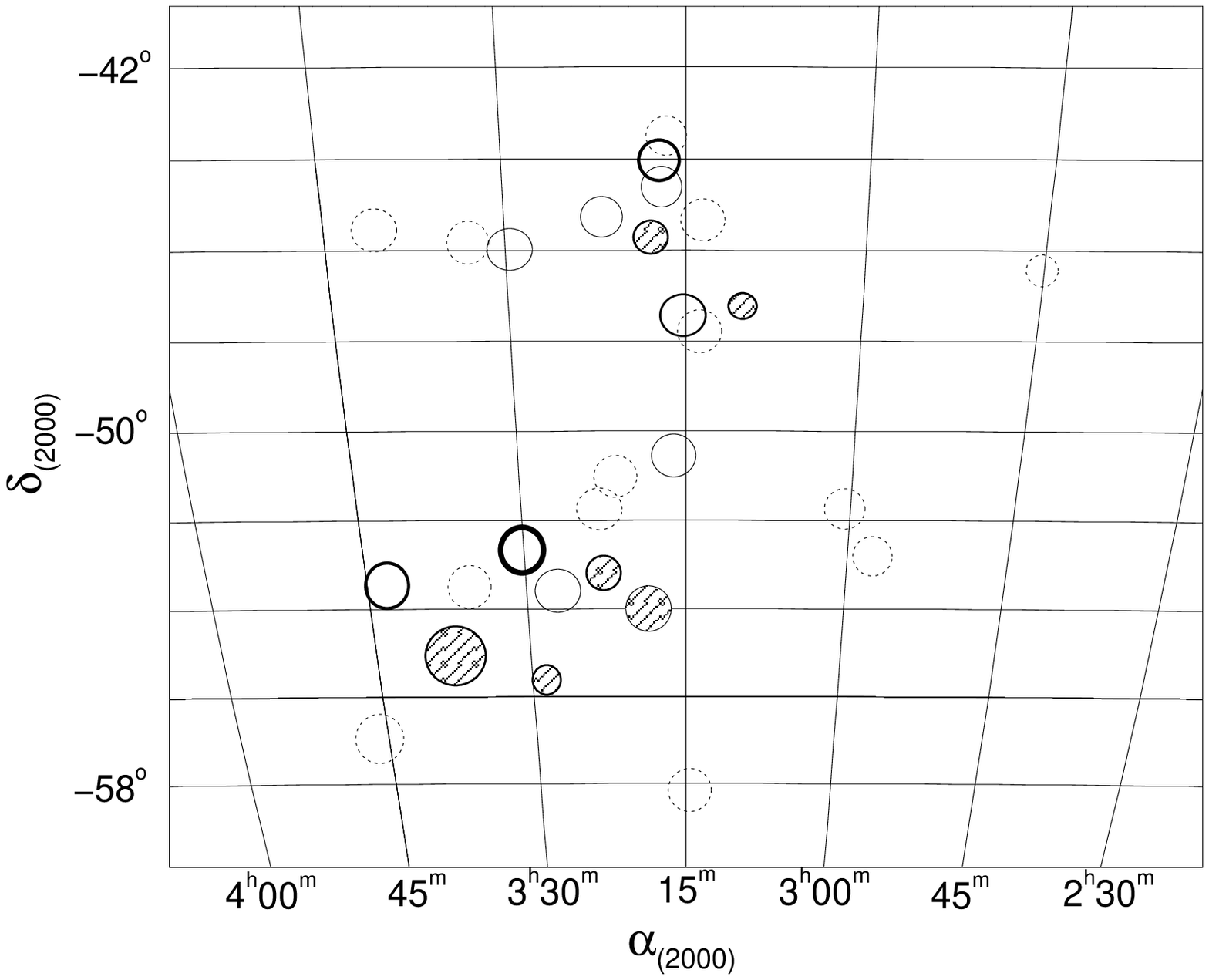}
  \caption{Hammer-Aitoff equal-area projection map of the HRS region,
  with galaxy clusters represented as circles. The radius of each
  cluster is scaled with the mean redshift to 0.5 Abell radii ($\sim$
  1 Mpc). The thickness of each outline represents the Abell Richness,
  where thicker lines are clusters of greater Richness Class. Dotted
  lines are for clusters with previously unreliable redshifts that
  have been improved in this study. Solid circles represent clusters
  whose mean redshifts from previous studies are reliable; these
  clusters are hatched if their redshifts fall outside the kinematic
  core of the HRS. \label{f1} } 
\end{figure}

Figure \ref{f1} shows the
spatial locations of the thirteen clusters in this study as dotted
circles. A further 15 clusters with secure redshifts, based on 10 or
more galaxies, are also displayed; those that fall within the
kinematic core of the HRS are shown as solid-line open circles.
Clusters that fall outside the statistically-defined kinematic core
may still, in fact, be members of the larger supercluster complex. Of the
thirteen clusters in the current study, eleven are ACO, Richness 0
clusters, and the remaining two are from the APMCC. Since the values
of cluster richness for the APMCC are not assigned in the same way as the
ACO, comparative determinations were taken from \cite{ein01} for the
two APMCC clusters, and they were found to be similar to ACO Richness 0.  

\section{OBSERVATIONS AND REDUCTIONS}
Spectroscopic observations were conducted 2004 November 14$-$17 with the
2.3m telescope of the Australian National University (ANU) at Siding Spring
Observatory. The Dual Beam Spectrograph (DBS) was utilized in
conjunction with a coated SiTE 1752$\times$532 CCD. The 300B
grating was used with all light directed into the blue 
arm via the insertion of a reflective mirror instead of the
customary dichroic. With a central wavelength of 5200 \AA, the above  
arrangement yielded a dispersion of 2.18 \AA\ pix$^{-1}$ from
[OII]$\lambda$3727 through Mg Ib$\lambda$5175 for the mean redshift of
the HRS. Wavelength calibration was based on CuAr lamp exposures carried
out after each object exposure. For each observation, the spectrograph
was rotated to place two or more galaxies on the slit.
Galaxies were selected based on their spatial proximity and their
apparent brightness. Specifically, all galaxies within a spatial
radius of $\leq$ 15$'$ ($= 0.5 R_{\rm{Abell}} \sim$1 Mpc) to the
published cluster center were examined and arranged in order of
decreasing brightness. We targeted only those galaxies with a blue,
$b_{\rm{J}}$,  magnitude brighter than 18.25, as given in the
SuperCOSMOS catalog \citep{ham01}. With a typical exposure time of 30
minutes, all spectra had signal-to-noise ratios of 15:1 or greater and
yielded accurate redshift determinations.   

Object exposures were reduced in the standard manner via the {\tt
IRAF}\footnotemark{} software package. Specifically, the following
steps were completed: debiasing, flat fielding, sky subtraction,
cosmic-ray removal, and wavelength calibration. Cosmic rays were
removed using the variance weighting option in the {\tt apall}
routine for aperture extraction. For those objects with
multiple exposures, the reduced spectra were co-added. In all, 76 usable
galaxy spectra were obtained over the four nights of observations, 
and they are listed in Table \ref{tb1} with their determined
redshift and associated uncertainty. Because the spectrograph position
angle was adjusted to allow two galaxies to be centered on the slit, observations
did not occur at the parallactic angle, and the uncertainty from
atmospheric dispersion could in principle be as much as 40
\kms. Radial velocities were determined for the galaxy spectra by
the standard technique of cross-correlating the galaxy spectra
against those of template stars. Stellar spectra of the G8III star HD
80499 and of the G4V star HD 106116 from the Indo-US Coud\'e Feed
Spectral library \citep{val04} and two de-redshifted DBS stellar
spectra (the G0 star HD 33771 and a serendipitous Galactic G dwarf at
$\alpha_{\rm {J2000}} =$ 03:29:38.44 and $\delta_{\rm {J2000}} =
-$52:36:08.5) were utilized as templates for the redshift determination 
using the {\tt xvsao} routine. Only cross-correlation fits with $R >$
4 \citep{ton79} were considered reliable and then averaged. For the
ten galaxies with emission-dominated features, procedures were
followed in a manner similar to that detailed previously in \po. As a
final step, all redshifts were corrected to the heliocentric reference
frame.  

\footnotetext{Image Reduction and Analysis Facility (IRAF) is written and
supported by the National Optical Astronomy Observatories (NOAO)
and the Association of Universities for Research in Astronomy (AURA),
Inc. under cooperative agreement with the National Science Foundation.} 

In addition to the new data from the ANU/DBS, 42 galaxy redshifts for
various clusters in our sample were obtained from other sources. Eighteen
cluster galaxies were observed during our survey with the multi-fiber,
6\degr\ field instrument \citep[6dF,][]{par98} on the UKST in 2004
November. Although that survey focused on the inter-cluster galaxies
in the HRS, otherwise unused fibers were placed on galaxies within the
clusters themselves. UKST/6dF spectra covered the wavelength range from
3900$-$7600 \AA\ and yielded average instrumental resolutions of 4.9
\AA\ and 6.6 \AA, for the 580V and 425R gratings respectively. The
automatic 6dF data reduction package completed the following:
debiasing, fiber extraction, cosmic-ray removal, flat-fielding, sky
subtraction, wavelength calibration, splicing, and co-addition
\citep{jon04}. The optical redshift for each galaxy was determined via
the semi-automated {\tt runz} software \citep{col01}, which employed
both cross-correlation for absorption features and emission-line
matching for typical features (e.g., [OII]$\lambda$3727,
[OIII]$\lambda$4959/5007, and Balmer lines). 

Furthermore, two previously unpublished datasets obtained with the
Anglo-Australian Telescope (AAT) were relied on for establishing
properties of certain clusters. Specifically, T. Mathams used the
fibre-optic-coupled aperture plate system \citep[FOCAP, see][]{gra83}
during 1986$-$1988 to observe galaxies within A3123 and APMCC 421 with
a dispersion of $\sim$ 2 \AA\ pix$^{-1}$ from 3600$-$5600
\AA. I. Klamer used the 2\degr\ field instrument \citep[2dF,][]{lew02} in
2002 January to observe galaxies within A3104 with $\sim$ 4 \AA\
pix$^{-1}$ (or 8 \AA\ FWHM) from 3600$-$8000 \AA. The overlap of 270
galaxies between the Mathams dataset and the observations from
\citet{ros02} revealed a velocity offset of 80 \kms\ within the Mathams
dataset. Therefore, a correction of $-$80 \kms\ was applied to all
redshifts cited by Mathams.   

The results of our observations, together with the other previously
unpublished data, are summarized in Table~\ref{tb1}.  The first column
contains the galaxy ID, while columns (2) and (3) list the J2000
coordinates, and column (4) gives the SuperCOSMOS $b_{\rm{J}}$ magnitude.  In
column (5) we give the velocity ($cz$) and
its associated uncertainty obtained from our ANU/DBS spectra. The
iterative method of calculating the mean cluster redshift and velocity
dispersion (described in \S4) shows that some galaxies are
either foreground or background to the cluster.  We label those galaxies
with an asterisk ($^{\ast}$) in column (5). Galaxy redshifts from the
literature (via the NASA Extragalactic Database, NED) are also utilized in our
calculations. All previously existing redshifts, either published or
unpublished, are listed in column (6) with their respective source in
column (7).

\section{DETERMINATION OF MEAN CLUSTER REDSHIFTS AND DISPERSIONS}

Given that galaxy clusters are thought to form via accretion along
intersecting filaments \citep[e.g.,][]{wes00}, and that such processes
are particularly pronounced in a dense environment like the HRS, we
expect the assumption of a gaussian velocity distribution for
the galaxies within the HRS clusters to be problematic. Specifically, the
probability of both projected and truly overlapping groups and/or
clusters will be enhanced within the supercluster environment.
Furthermore, the calculation of the cluster mean redshift and velocity
dispersion under an assumption of gaussian statistics is neither
robust nor efficient \citep{pea31,box53}. \citet[][hereafter
  BFG90]{bee90} define a number of 
reliable estimators for the mean cluster redshift (location) and
dispersion (scale) that are more robust to the presence of outliers
and less wed to the gaussian assumption. For a small number of galaxy
redshifts per cluster (N$_{\rm{gx}} < 20$), we utilize the biweight
estimator for calculating both the location ($C_{\rm{BI}}$) and the scale
($S_{\rm{BI}}$) of each cluster according to the following:
\begin{equation}
C_{\rm{BI}} = M + \, \frac{\sum_{\rm{\vert u_i \vert < 1}} \, (x_{\rm{i}} - M)(1 -
  u_{\rm{i}}^2)^2}{\sum_{\rm{\vert u_i \vert < 1}} \, (1 - u_{\rm{i}}^2)^2},
\end{equation}
where $M$ is the sample median and $u_{\rm{i}}$ are the individual
weights as defined by:
\begin{equation}
u_{\rm{i}} = \frac{(x_{\rm{i}} - M)}{c(\rm{MAD})}.
\end{equation} 
The tuning constant, $c$, establishes the low and high velocity cutoff
for each cluster. The median absolute deviation, \rm{MAD}, is defined by: \rm{MAD}
$=$ median$(\vert x_{\rm{i}} - M \vert)$. Improvements are made in the
final location (and scale) of the cluster by iteratively substituting
the most recently calculated $C_{\rm{BI}}$ for the value of $M$, and then
re-calculating a new $C_{\rm{BI}}$ until convergence is achieved
(BFG90).  Although we experimented with different values of $c$
to evaluate the sensitivity of the results on that parameter, the $c$
parameter was held at the suggested value of 6.0, which excludes all
data that are more than 4 standard deviations from the central
location.  While $c$ was varied from 4.0 $-$ 10.0 for the $C_{\rm{BI}}$
parameter, the maximum change observed for each cluster remained
within the estimated uncertainty of $C_{\rm{BI}}$ in Table \ref{tb2}
(column 8). 

In a similar way, the biweight estimator for scale, $S_{\rm{BI}}$, is given
by:
\begin{equation}
S_{\rm{BI}} = n^{1/2} \, \frac{[\sum_{\rm{\vert u_i \vert < 1}} \,
  (x_{\rm{i}} - M)^2(1 - 
  u_{\rm{i}}^2)^4]^{1/2}}{\vert \sum_{\rm{\vert
  u_{\rm{i}} \vert < 1}} \, (1 - u_{\rm{i}}^2)(1 - 5u_{\rm{i}}^2) \vert},
\end{equation}
with the same definitions as above only here, as suggested, $c$ was
set to 9.0. Again, the routine was iterated until convergence. Moreover, 
varying $c$ from 5.0 $-$ 11.0 resulted in a typical total scale
change of only $\Delta S_{\rm{BI}} \leqslant$ 50 \kms. Although BFG90
adopt the terminology of ``location'' and ``scale'' because of the 
difference in definition between these parameters and the canonical mean
and dispersion, we retain the common usage of the cluster mean
redshift and velocity dispersion for the rest of the paper. 

Data on the cluster mean redshifts (location) and velocity dispersions (scale)
are summarized in Table~\ref{tb2}.  The previously published value for the 
mean redshift is given in column (5), with the source for that redshift in
column (6).  In column (7) we list the number of galaxies (i.e., those
from Table \ref{tb1}, columns (5) and (6), excluding
foreground and background galaxies) on which our new mean redshift is
based. The new cluster redshift and associated uncertainty are
given in column (8). Finally, the newly determined velocity dispersion
is given in column (9).  For the three cases in which the cluster
appears to consist of multiple components, and thus has a less-reliable
mean redshift and $\sigma$, we have followed the values in column (8)
and (9) with a colon (:).  These three special cases are discussed
further in \S\ 5. 

For the remaining ten clusters in the study, the new observations
provide a sufficient increase in the number of known redshifts to
allow us to determine a reasonably secure velocity
dispersion. Furthermore, all of the clusters are Abell richness class,
$R = 0$ (or APMCC equivalent), hence we can assess the mean and
scatter in velocity dispersion for $R =$ 0 clusters in the HRS. Given
the modest size of our sample of cluster velocity dispersions, we
utilize the same routine for the biweight location estimator,
$C_{\rm{BI}}$, to determine an effective mean velocity dispersion for
our cluster sample. After excluding three values as outliers, the remaining
ten clusters give a mean velocity dispersion of 420 $\pm$ 50 \kms\
for Richness 0 clusters. This result is intermediate between published
values for galaxy groups (both loose at 165 \kms\ in \citet{tuc00} and
compact at $\sim$ 250 \kms\ in \citet{hic97}) and rich galaxy clusters
(i.e., larger structures) at $\sim$ 700 \kms\ \citep{maz96}.  

\section{RESULTS FOR INDIVIDUAL CLUSTERS}

The new observational data in three clusters result in velocity dispersions
that are quite large in comparison with the $\sim$400 \kms\ mean value for
Richness 0 clusters found above. Although \citet{maz96} find a large
intrinsic scatter in velocity dispersion for rich clusters,
the derived dispersions for these three clusters rival (and exceed)
the $upper$ limits observed by the same authors for R $\geqslant$ 1
clusters. Therefore, the presence of multiple components and/or spatial
projection of multiple clusters/groups is suggested. We examine these
three systems in greater detail, since their true composition remains
unclear.

\subsection{Abell 3047/ APMCC 290 (02\h\ 45\fm 25 \hash 46\ad\ 26\farcm0)} 

The structure of this R $ = 0$, D $= 6$ cluster
is quite regular in shape and centers around the brightest cluster
galaxy (BCG), 2MASX J02451334$-$4627194 ($b_{\rm{J}} =$ 16.68), whose
previously published redshift is 27,581 \kms\ \citep{gra02}. We
observe a redshift of 28,279 $\pm$ 65 \kms\ for the same galaxy,
where the difference is most likely due to the higher resolution of
the DBS spectra (4.5 \AA\ compared to 15$-$20 \AA). This result is
consistent with the mean cluster redshift given in SR99 of 0.0950
(28,500 \kms), which is based on fewer than 4 galaxy redshifts
(``N$_{\rm{gx}} <$4''). X-ray emission is also detected at a level of
$L_{\rm{x}} = 3.86\times\ 10^{43}\,$ergs s$^{-1}$ \citep{cru02} and is
centered on the BCG, thereby strengthening the idea that at least one
significant cluster is present.  

The iterative biweight estimator routine does not
exclude any of the 8 proposed members, and the following results are
obtained: $\overline{cz} =$ 27,382 \kms\ and $\sigma =$ 1225 \kms. While
the mean redshift of the cluster is somewhat similar to the previous
result for the BCG, the derived velocity dispersion is too inflated
for a cluster of Richness 0. In seeking an
alternative explanation, we notice that three of the four brightest
galaxies have a noticeably different recessional velocity ($\leqslant$
2000 \kms) than the majority. Therefore, we may be viewing the
projection of two separate systems, or a physical overlap/merger,
giving the appearance of a single R $=$ 0 cluster. By subdividing out
the galaxies in the following way, a more logical result is obtained: \\   
\noindent C1:  N$=$3, $\overline{cz} =$ 26,285 \kms,
$\sigma =$ 620 \kms;  C2:  N$=$5,  $\overline{cz} =$ 28,275 \kms,
$\sigma =$ 725 \kms.\\
We note that the dispersions for the two components are still
excessive for a Richness 0 cluster. On the other hand, the presence of
X-ray emission at the observed level is consistent with a $R =$
1 or 2 cluster \citep[][Figure 9]{led03}, which lends support to the
high velocity dispersions found for the two components. Furthermore,
such a large value of $L_{\rm{x}}$ is also consistent with a cluster
merger along the line of sight. In any case, the redshifts of both
components are well outside the kinematic core of the HRS. 

\subsection{Abell 3109 (03\h\ 16\fm 5 \hash 43\ad\ 51\farcm 0)}

Although we add no new observations in this cluster, a compilation
of 14 previously published galaxy redshifts provides more established kinematic
properties. The ESO Nearby Abell Cluster Survey \citep[ENACS,][]{kat98}
focused on rich clusters with R $\geqslant$ 1. The periphery of
Abell 3112 \citep[03\h\ 17\fm 9 $-$44\degr\ 14\farcm 0, R $=$ 2, $cz =
  $ 22,500 in][]{maz96} overlaps with A3109, providing us with 9 
redshifts from the ENACS data. The assumed BCG in A3109, 2MASX
J03163934$-$4351169, $b_{\rm{J}} =$ 15.60, has a published redshift of 18,594
\kms\ \citep{mur95}, which is inconsistent with the published value for
the cluster (27,581 \kms\ in SR99, see their note). 

By incorporating all galaxy redshifts within the prescribed radius,
the biweight estimator selects 11 cluster members with the following
kinematic properties:  $\overline{cz} =$ 18,950 \kms\ and $\sigma =$ 850
\kms. Reducing the radial extent to 13$'$ and thereby excluding 2
proposed members, we obtain a slightly
decreased dispersion: $\overline{cz} =$ 18,850 \kms and $\sigma =$ 700
\kms. Even though the dispersion remains greater than the $\sim$400
\kms\ mean for R $=$ 0 clusters that we obtained earlier, the
archived redshift information establishes a reliable cluster
location (i.e., mean redshift) and places A3109 within the HRS. 

\subsection{Abell 3120 (03\h\ 22\fm 0 \hash 51\ad\ 19\farcm 0)} 

The \rich 0, \dist 5 cluster, A3120, for which we have            
obtained 5 galaxy redshifts, is the nearest cluster to the
published spatial center of the HRS \citep{zuc93}. Its published
redshift of 20,700 \kms\ (SR99) is also close to the $\sim$19,900
\kms\ mean redshift of the HRS (\po). While A3120 does not meet the
specific cluster criteria for the APMCC, it does contain the bright galaxy,
2MASX J03215645$-$5119357 ($b_{\rm{J}} =$ 15.91), with a previously 
published redshift of 21,040 \kms\ \citep{luc83}. The biweight
estimator routine accepts all observed galaxies, and we derive
the following cluster properties: $\overline{cz} =$ 20,525 \kms, $\sigma
=$ 1400 \kms. While the mean derived velocity is in accord with the
published value, the large dispersion is clearly inconsistent with an
\rich 0 cluster.  Furthermore, the five galaxies with redshift
information show no discernible spatial or kinematic segregation, as
one might expect with a cluster. Hence there is reason to suspect that
A3120 is not a cluster but the projection of many inter-cluster galaxies
near the center of the HRS. 

On the other hand, \citet{rom00} find X-ray emission at a level of
$L_{\rm{x}} = 2.22\times\ 10^{43}\, $ergs s$^{-1}$, centered on 2MASX
J03215645$-$5119357, and propose that the X-rays are emitted by a
``fossil group'' (see their Figure 20). These groups form as a result
of multiple mergers within the group or a cluster that lead to a
single dominant giant elliptical surrounded by an X-ray halo
\citep{pon94,jon03}. However, the X-ray position also coincides with
a radio source from the Sydney University Molonglo Sky Survey (SUMSS),
SUMSS J032156$-$511935, with a flux density of 49.0 mJy \citep{mau03}.
Considering the wide range of X-ray luminosities in active galactic
nuclei (AGN), it is conceivable that some (or all) of the X-ray emission
is a result of the AGN, rather than the fossil halo. Because the
galaxy's redshift is taken from the literature and no optical spectrum
is available, we conclude that the situation in A3120 is not soluble
with the current observational data. In Table \ref{tb2} we give the
formal mean redshift (location), uncertainty, and dispersion (scale) as
deduced from the biweight estimator analysis. However, since we believe that
the most likely value of the actual cluster redshift is that of the
(presumed) BCG 2MASX J03215645$-$5119357 ($cz =$ 21,040 \kms), we adopt this
value for the mean redshift of A3120 in Table \ref{tb3} (noted by the
``1'' in column 7). Fortunately, the difference in redshift between
20,700 \kms\ in Table \ref{tb2} and 21,040 \kms\ in Table \ref{tb3}
is within the biweight uncertainty. 

\section{REDSHIFT DISTRIBUTION OF THE HRS CLUSTERS}
\subsection{Consistency with the Inter-cluster Galaxies}

\begin{figure}
  \plotone{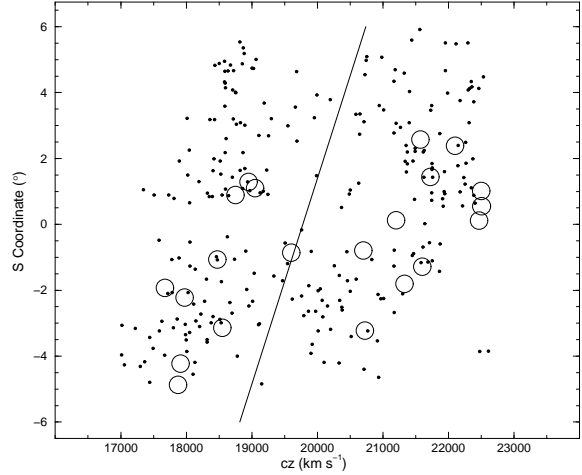}
  \caption{Projected angular $S$-coordinate is plotted versus redshift
  for 6dF inter-cluster galaxies from \po\ (small filled circles)
  between 17,000 
  and 22,500 \kms\ at a PA $= -$80\degr. Open circles represent
  the location and approximate extent of the clusters in the region
  with mean redshifts listed in Table \ref{tb3}. The solid line is the
  best fit to the inter-cluster data at this PA. \label{f2} }  
\end{figure}

\begin{figure}
  \plottwo{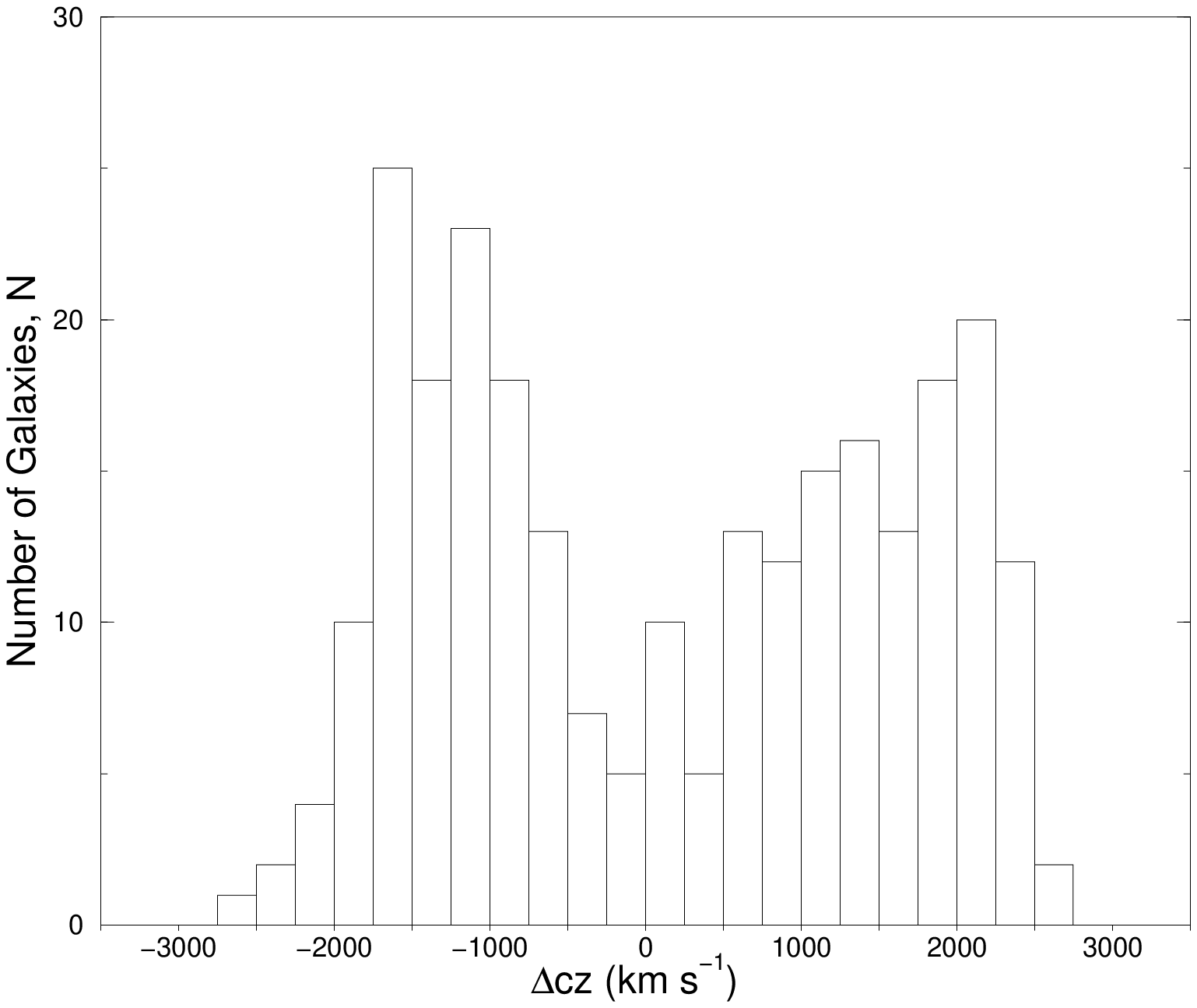}{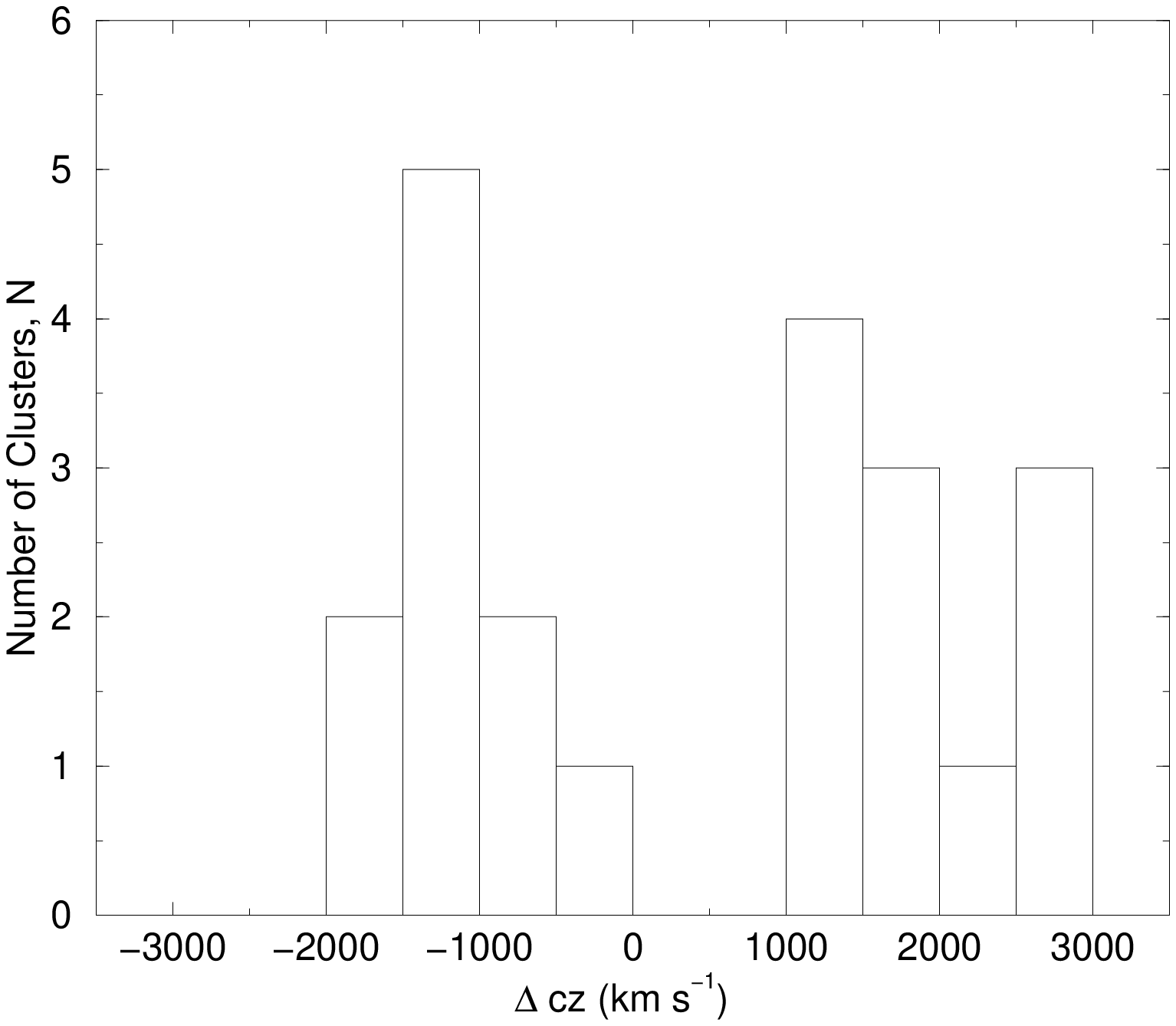}
  \caption{Histograms of residual redshifts along the best-fit
  line at a PA $= -$80\ad , shown as the solid line in Figure
  \ref{f2}. \textbf{Left}: Previous results for the HRS inter-cluster
  galaxies; \textbf{Right}: New results for the HRS clusters as listed in Table
  \ref{tb3}.\label{f3} } 
\end{figure}

In \po, the inter-cluster galaxies within the range 17,000 $-$ 22,500 \kms\
(i.e., the HRS kinematic core) were found to exhibit a systematic
$\sim$1500 \kms\ increase in redshift with position along a southeast-northwest
axis. To quantify this spatial-redshift correlation, we projected
the inter-cluster galaxies in the HRS onto an (assumed) principal
axis through the spatial center of the HRS. The projected distance
along the principal axis was referred to as the $S$-coordinate. We then
performed a linear correlation analysis between the redshift and the
$S$-coordinate. After varying the position angle (PA) of the
principal axis over the full range of PA and repeating the correlation
analysis at each PA, we found that the inter-cluster galaxies show the
highest correlation coefficient ($R =$ 0.3) and lowest 
probability for no correlation ($P <$ 10$^{-6}$) at a PA $\approx$
$-80$\arcdeg\ (as measured east from north). Further details are
provided in \po. In Figure \ref{f2}, we plot the projected $S$
position versus redshift for all inter-cluster galaxies and clusters with
redshift between 17,000 and 22,500 \kms\ at a PA of $-80$\arcdeg. As
is discussed in \po, the distribution of redshifts in
Figure \ref{f2} is clearly divided into two main components; one
centered at $\sim$18,000 \kms\ and the other at $\sim$21,000 \kms. Furthermore,
there is a correlation between the $S$-coordinate and redshift for
both of the components in the sense that the redshift increases
systematically from the southeast (negative $S$-coordinate) to the
northwest (positive $S$-coordinate). Once the kinematic trend is
accounted for and removed, the bi-modal nature of the redshift
distribution of the inter-cluster galaxies becomes even more
apparent. We have plotted the histogram of residual redshifts (after
removal of the overall kinematic trend) for the inter-cluster galaxies
in the left panel of Figure \ref{f3}. Using the KMM statistical
test \citep{ash94}, we find that a two-gaussian fit to the redshift
histogram is preferred to a single gaussian at a confidence level of
$>$99.9\%. 

Although the published cluster redshifts in \po\ followed the
spatial-redshift trend of the inter-cluster galaxies, we found
that the histogram of residual redshifts showed no evidence for
bi-modality. This fact appeared somewhat puzzling given the
expectation that inter-cluster galaxies and clusters in the same area
of the sky should have similar redshift distributions. However, the
new cluster redshift data compiled in Table \ref{tb3} shows a different
signature. Plotted as large open circles in Figure \ref{f2}, the
HRS clusters with reliable redshifts now also appear to divide into two main 
components. Furthermore, this impression is supported by the histogram
of cluster residual redshifts, plotted in the right hand panel of Figure
\ref{f3}. Again, the KMM statistical test is applied to the
histogram, and we find that a two-gaussian fit is favored over a
single gaussian at the 99.8\% confidence level. Moreover, the
separation of $\sim$3000 \kms\ between the two peaks of the cluster
redshift distribution is consistent with the similar figure found
between the two peaks in the inter-cluster galaxy redshifts. Upon
implementing the spatial-redshift correlation analysis described above
(i.e., projection of the cluster positional data onto a
principal axis), we find a correlation coefficient of $R = $ 0.5 and
a probability of no correlation of 10$^{-2}$ at a PA of $-80$\degr\
for the principal axis. In summary, with the improved mean
redshift data for many of the clusters in the HRS, we conclude that
the overall redshift distributions of the clusters and the 
inter-cluster galaxies are now consistent with each other. A closer
examination of the inter-relationship between clusters and 
inter-cluster galaxies in the HRS (e.g., an evaluation as to whether
the clusters are indeed located at the intersection of galaxy
filaments) awaits a more comprehensive dataset that is in progress.

\subsection{Re-determination of the Kinematic Core}

As mentioned in \S 2 and above, we determined rough kinematic
boundaries in \po\ for the the HRS complex, referred to as the
kinematic core, by fitting a gaussian to the redshift distribution of the 
Abell clusters listed as HRS members by both \citet{zuc93} and
\citet{ein94}. We used the mean redshift and the FWHM of the
distribution to define the kinematic core. Given that we now have
reliable mean redshifts for 16 of these 17 Abell clusters (we exclude
A3120 for reasons discussed in \S 5.3), it is worth investigating
whether the kinematic core changes significantly as a result of the
improved redshift data. Using column (6) in Table
\ref{tb3} for the 15 Abell clusters (``A'' designation) plus the
redshift for A3093 \citep[$\overline{cz} =$ 24,900 \kms,][]{kat98},
the following values for the mean (location) and the dispersion (scale) are
obtained by utilizing the biweight estimator: $\overline{cz} =$ 20,150
$\pm$ 525 \kms\ and $\sigma =$ 2125 \kms. These values imply that the
kinematic core of the HRS lies between 17,700 and 22,700 \kms. Hence
the kinematic center of the HRS is slightly higher than the previous
value, and the core is slightly narrower. 

In light of the above discussion, none of our previous results are
significantly altered if we use these revised values to define which
clusters should be included in the redshift bi-modality
analysis. Furthermore, the inter-cluster galaxies from \po\ continue
to show a preferred spatial-redshift axis 
at PA $= -80\degr$ with similar correlation values over the
somewhat-modified range of 17,700 $-$ 22,700 \kms. Moreover, our
definition of the kinematic core is only suggestive of what should be
included in a detailed analysis of the complex structure of the
HRS. Clearly, the actual boundaries of the HRS can be expected to
extend to some clusters and inter-cluster galaxies outside the
immediate kinematic core.  

\subsection{Comparisons with the Shapley Supercluster}

Finally, we compare the improved cluster redshift distribution for the
HRS with that of the Shapley supercluster \citep[hereafter SSC,][]
{qui95,qui00}. While the 23 Abell clusters in the SSC for which
reliable redshift information is published definitely show a bi-modal
distribution, there is a 3:1 number imbalance between the two cluster
redshift peaks in the SSC. That is, there are many more
clusters in the higher redshift peak, which contains the most massive
cluster in the complex, Abell 3558 ($\overline{cz} = $14,500 \kms),
than there are in the lower redshift peak at $\sim$ 11,000 \kms. In
contrast, the HRS clusters are equally split between the two redshift peaks as
determined from Figure \ref{f3}. The difference in mean redshift
for the two peaks is slightly higher for the SSC ($\sim$3500 \kms) as
opposed to the $\sim$3000 \kms\ difference in the HRS. Furthermore, as
discussed in \po, 
the redshift distribution for the inter-cluster galaxies in both the
HRS and SSC are bi-modal, with a roughly equal split between the two
redshift peaks for both clusters \citep{dri04}. In short, while there
are striking similarities between the two largest mass concentrations
in the local universe, the 3:1 imbalance in the number of clusters in
the redshift peaks of the SSC represents an interesting contrast with
the more evenly distributed HRS.

\section{CONCLUSIONS}

We have obtained 76 new optical redshifts within 12 galaxy clusters of the
Horologium-Reticulum supercluster (HRS). These observations, augmented
by 42 previously unpublished redshifts, have led to the
determination of more accurate cluster properties. Using the methods
for calculating robust mean redshifts (location) and velocity
dispersions (scale) described in BFG90, we have calculated mean
redshifts and dispersions for 13 clusters, including A3109 for which
no new observations are reported. The mean redshifts for several
clusters have changed by at least 750 \kms\ (in 6/13 observed) from
their previously reported values. In addition, three clusters are observed
to consist of multiple components (A3047, A3109, and A3120). The new cluster
redshift data have been compared to previously compiled redshift data
for the inter-cluster galaxies in the HRS from \citet{fle05}. Primarily, 
we now find consistency between the large-scale kinematic features of the
clusters and the inter-cluster galaxies. Specifically, there is a
principal kinematic axis in the HRS at a PA of $-$80\degr\ east from
north, along which a systematic increase in redshift with position is
observed for both clusters and inter-cluster galaxies. After this
overall spatial-kinematic trend is removed, the distribution in
redshift for both clusters and inter-cluster galaxies is distinctly
bi-modal, with the two redshift peaks separated by $\sim$3000 \kms.

We thank the Australian National University and Mount Stromlo/Siding
Spring Observatories for facilitating and supporting these
observations. We also thank Clair Murrowood for her assistance with
the observations, Ilana Klamer for supplying her unpublished
2dF data of A3104, and Bruce Peterson for the use of Mathams' thesis
data. M. C. F. acknowledges the support of a NASA Space
Grant Graduate Fellowship at the University of North Carolina-Chapel
Hill. R. W. H. acknowledges grant support from the Australian Research
Council. M. J. H. acknowledges support through IRGS Grant J0014369
administered by the University of Tasmania. A
portion of this work was supported  by NSF grants AST-9900720 and
AST-0406443 to the University of North Carolina-Chapel Hill. This
research has made use of the NASA/IPAC Extragalactic Database 
(NED) which is operated by the Jet Propulsion Laboratory, California
Institute of Technology, under contract with the National Aeronautics
and Space Administration. 

\bibliographystyle{apj}
\bibliography{mcfrefs}

\newpage
\begin{deluxetable}{ccccccc}
\tabletypesize{\footnotesize}
\tablecaption{ Redshift Data for Galaxy Clusters in Horologium-Reticulum
  \label{tb1}} 
\tablewidth{0pt}
\tablehead{
\colhead{IAU Name} & \colhead{$\alpha_{\rm{2000}}$}   & \colhead{$\delta_{\rm{2000}}$} &
\colhead{$b_{\rm{J}}$} & 
\colhead{$cz \pm u _{\rm{cz}}$ (\kms) }  & \colhead{$cz_{\rm{pub}}$ (\kms)}
  &  \colhead{Source}\\
\colhead{(1)} & \colhead{(2)}   & \colhead{(3)} &
\colhead{(4)}  & \colhead{(5)} & \colhead{(6)} &
\colhead{(7)} 
} 
\startdata
 & & Abell 3047 & & &\\
\hline
2MASX J02445221$-$4630015 &02 44 52.22  &$-$46 30 01.3 & 17.15& 26252 $\pm$
63& 26355 & 1\\
2MASX J02450315$-$4628464 &02 45 03.10  &$-$46 28 46.2 & 18.24 &	28108
$\pm$ 72& \\  
2MASX J02450401$-$4626435&02 45 03.98  &$-$46 26 43.6 & 17.56& 26921 $\pm$ 72& \\ 
2MASX J02450895$-$4626245&02 45 08.96  &$-$46 26 24.9 & 18.24 &27396 $\pm$ 53& \\ 
2MASX J02451207$-$4628013&02 45 12.14  &$-$46 28 00.9 & 17.97&28181 $\pm$ 82& \\ 
..... & ..... & ..... & ..... & ..... & ..... & .. \\ 

\enddata

\tablecomments{Numbers in parentheses apply to column numbers. (1) IAU
  Name; (2) Right Ascension (J2000); (3) Declination (J2000); (4)
  SuperCOSMOS $b_{\rm{J}}$ apparent magnitude; (5) radial velocity,
  $cz$, with associated uncertainty; (6) Previously published redshift;
  (7) Source of published redshift: 1$-$ 6dF observations, unpublished
  (2004),...  Table \ref{tb1} is published in its entirety in the
  electronic edition of the {\it Astronomical Journal}. A portion is
  shown here for guidance regarding its form and content.}
\end{deluxetable}

\begin{deluxetable}{ccccccccc}
\tabletypesize{\scriptsize}
\tablecaption{ Revised Mean Redshifts and Velocity Dispersions for HRS Clusters
  \label{tb2}} 
\tablewidth{0pt}
\tablehead{
\colhead{Cluster} & \colhead{$\alpha_{\rm{2000}}$}   & \colhead{$\delta_{\rm{2000}}$} &
\colhead{N$_{\rm{gx, prev}}$}  & \colhead{cz$_{\rm{prev}}$ (\kms)} & \colhead{Source} &
\colhead{N$_{\rm{gx, new}}$}  & \colhead{$\overline{cz}_{\rm{obs}} \pm
  u_{\rm{\bar{cz}}}$ (\kms)} & \colhead{$\sigma$ (\kms)} \\
\colhead{(1)} & \colhead{(2)}   & \colhead{(3)} &
\colhead{(4)}  & \colhead{(5)} & \colhead{(6)} &
\colhead{(7)}  & \colhead{(8)}
 & \colhead{(9)}
}
\startdata
A3047 & 02 45.2 &$-$46 27.0 & $<4$ & 28500 & 1 & 8 & 27550: $\pm$ 425 &
1225:\\
A3074 & 02 57.9 &$-$52 43.0 & $<4$  & 21900 & 1 & 7 & 21575 $\pm$ 125 & 325\\
A3078 & 03 00.5 &$-$51 50.0 & $>0$& 19440 & 1 & 8    & 22100 $\pm$ 200
& 575\\
A3100 &	03 13.8	&$-$47 47.0 & $>0$& 18870 & 1 & 9    & 19050 $\pm$ 75 & 250\\
A3104 &	03 14.3 &$-$45 24.0 & $<4$  & 21885 & 1 & 28   & 21725 $\pm$ 125 &
700\\
A3106 & 03 14.5 &$-$58 05.0 & $>0$& 19170 & 1 & 7    & 19600 $\pm$ 115 & 300\\
A3109 & 03 16.6 &$-$43 51.0 & 1 & 27240 & 2 & 11   & 18950: $\pm$ 250 &
850: \\
A3120 & 03 21.9 &$-$51 19.0 & $>0$& 20700 & 1 & 5    & 20525: $\pm$ 675
& 1400:\\ 
A3123 & 03 23.0 &$-$52 01.0 & $>0$& 19320 & 1 & 11    & 18475 $\pm$
100 &  375\\ 
A3133 &	03 32.7 &$-$45 56.0 & $>0$& 16290 & 1 & 7    & 21325 $\pm$ 175 & 475\\
APMCC 421 & 03 35.5& $-$53 40.9 & 2 & 18887 & 2 & 11  & 18550 $\pm$
100 & 300\\  
APMCC 433 & 03 41.1& $-$45 41.5 & 2& 19786 & 2 & 11  & 20725 $\pm$ 125& 425\\ 
A3164 &	03 45.8 &$-$57 02.0 & 3& 17100 & 1 & 7 & 17875 $\pm$ 225 & 575\\
\enddata

\tablecomments{Numbers in parentheses apply to column numbers. (1)
  Cluster name; (2) Right Ascension in hours and
  minutes (J2000); (3) Declination in degrees and minutes (J2000); (4)
  Number of galaxies used to establish previously published mean
  redshift, where ``N$_{\rm{gx}}>0$'' and ``N$_{\rm{gx}}<4$'' are
  designations given by SR99 to reflect the ambiguity regarding 
  the number of individual velocities from the original source; (5) previously
  published mean redshift; (6) Source for published mean redshift: 1
  $-$ SR99 and 2 $-$ \citet{dal94,dal97}; (7) Number of galaxies on which
  the new cluster properties were based; (8) new mean cluster redshift
  and associated uncertainty; (9) new cluster velocity dispersion.}
\end{deluxetable}

\begin{deluxetable}{cccccccc}
\tabletypesize{\scriptsize}
\tablecaption{ Reliable Cluster Redshifts in the HRS Kinematic Core
  \label{tb3}} 
\tablewidth{0pt}
\tablehead{
\colhead{Cluster} & \colhead{$\alpha_{\rm{2000}}$}   &
\colhead{$\delta_{\rm{2000}}$} & 
\colhead{Richness} &
\colhead{Redshift, $\overline{z}$}  & \colhead{$\overline{cz}$ (\kms)}
& \colhead{N$_{\rm{gx}}$} & \colhead{Source} \\
\colhead{(1)} & \colhead{(2)}   & \colhead{(3)} &
\colhead{(4)}  & \colhead{(5)} & \colhead{(6)} &
\colhead{(7)} & \colhead{(8)}
}
\startdata
A3074 & 02 57.9 & $-$52 43.0 & 0 & 0.071917 & 21575 & 7  & 1\\
A3078 & 03 00.5 & $-$51 50.0 & 0 & 0.073767 & 22100 & 8  & 1\\
A3100 &	03 13.8	& $-$47 47.0 & 0 & 0.063500 & 19050 & 9  & 1\\
A3104 &	03 14.3 & $-$45 24.0 & 0 & 0.072417 & 21725 & 28 & 1\\
A3106 & 03 14.5 & $-$58 05.0 & 0 & 0.065333 & 19600 & 7  & 1\\
A3108 & 03 15.2 & $-$47 37.0 & 1 & 0.062500 & 18750 & 7  & 2\\
A3109 & 03 16.7 & $-$43 51.0 & 0 & 0.063167 & 18950 & 11 & 1\\
A3110 &	03 16.5	& $-$50 54.0 & 0 & 0.074900 & 22470 & 10 & 3\\
APMCC 369 & 03 17.5 & $-$44 38.5 & 0 & 0.075000 & 22500 & 29 & 3\\
A3112 & 03 17.9 & $-$44 14.0 & 2 & 0.075000 & 22500 & 77 & 2\\
S0345 & 03 21.8 & $-$45 32.3 & 0 & 0.070667 & 21200 & 18 & 4\\
A3120 & 03 21.9 & $-$51 19.0 & 0 & 0.070133 & 21040 & 1  & 5\\
A3123 & 03 23.0 & $-$52 01.0 & 0 & 0.061583 & 18475 & 11 & 1\\
A3125 & 03 27.4 & $-$53 30.0 & 0 & 0.058900 & 17670 & 40 & 6\\
S0356 & 03 29.6 & $-$45 58.8 & 0 & 0.072000 & 21600 & 8  & 3\\
A3128 &	03 30.2	& $-$52 33.0 & 3 & 0.059900 & 17970 & 158 & 2\\
A3133 &	03 32.7 & $-$45 56.0 & 0 & 0.071083 & 21325 & 7  & 1\\
APMCC 421& 03 35.5& $-$53 40.9 & 0& 0.061833 & 18550 & 11 & 1\\ 
APMCC 433& 03 41.1& $-$45 41.5 & 0-1 & 0.069083 & 20725 & 11 & 1\\ 
A3158 &	03 43.0 & $-$53 38.0 & 2 & 0.059700 & 17910 & 105 & 2\\
A3164 &	03 45.8 & $-$57 02.0 & 0 & 0.059583 & 17875 & 7  & 1\\
\enddata

\tablecomments{Numbers in parentheses apply to column numbers. (1)
  Cluster name, where ``S'' denotes poor clusters from ACO; (2) Right
  Ascension in hours and minutes (J2000); (3) Declination in degrees and 
  minutes (J2000); (4) ACO Richness or APMCC equivalent; (5)
  Average redshift taken from Source; (6) Recessional velocity; (7)
  Number of galaxies used to calculate kinematic properties, for A3120
  see \S5.3; (8) 1$-$ This study, 2$-$ \citet[ENACS,][]{kat98}, 3$-$
  \citet{alo99}, 4$-$ I. Klamer (private communication), 5$-$
  \citet{luc83}, 6$-$\citet{cal97}. }   
\end{deluxetable}

\end{document}